\documentstyle[aps,psfig,epsf]{revtex}

\begin{document}
\title{Resistance of a domain wall in the quasiclassical approach}
\author{F.S.Bergeret $^{1 }$, A.F. Volkov$^{1,2}$ and K.B.Efetov$^{1,3}$}
\address{$^{(1)}$Theoretische Physik III,\\
Ruhr-Universit\"{a}t Bochum, D-44780 Bochum, Germany\\
$^{(2)}$Institute of Radioengineering and Electronics of the Russian Academy%
\\
of Sciences, 103907 Moscow, Russia \\
$^{(3)}$L.D. Landau Institute for Theoretical Physics, 117940 Moscow, Russia }
\maketitle

\begin{abstract}
Starting from a simple microscopic model, we have derived a kinetic equation
for the matrix distribution function. We employed this equation to calculate
the conductance $G$ in a mesoscopic F'/F/F' structure with a domain wall
(DW). In the limit of a small exchange energy $J$ and an
abrupt DW, the conductance of the structure is equal to $G_{2d}=4\sigma
_{\uparrow }\sigma _{\downarrow }/(\sigma _{\uparrow }+\sigma _{\downarrow
})L$. Assuming that the scattering times for electrons with up and down spins
are close to each other we show that the account for a finite width of the DW leads to an increase in this
conductance. We have also calculated the spatial distribution of the
electric field in the F wire. In the opposite limit of large $J$ (adiabatic
variation of the magnetization in the DW) the conductance coincides in the
main approximation with the conductance of a single domain structure $%
G_{1d}=(\sigma _{\uparrow }+\sigma _{\downarrow })/L$. The account for
rotation of the magnetization in the DW leads to a negative correction to
this conductance. Our results differ from the results in papers published
earlier.
\end{abstract}

\section{Introduction}

In ferromagnetic metals not only the charge of the electron but also the
spin plays an important role in transport phenomena. A famous example is the
observation of the giant magnetoresistance in magnetic multilayers, which
can be explained in terms of a spin dependent electronic scattering.

The presence of a domain wall (DW) in a ferromagnet can also change transport
properties and this has been observed in a number of experiments. At the
first glance, experimental data seems to contradict to each other{. In Refs. 
\cite{gregg,ebels,ruediger_prb,viret} it was found that the resistance of
ferromagnetic wires and films decreases when increasing the external
magnetic field, whereas in Refs. \cite{taniyama,ruediger} the resistance at
zero magnetic field was found to be smaller than the one measured at high
magnetic fields. }

{In order to give a quantitative description of these experiments not only
the DW contribution to the magnetoresistance (MR) should be taken into
account but also other mechanisms, as the anisotropic magnetoresistance,
which arises due to the spin-orbit scattering \cite{smit,berger,potter},
size effects and the Lorentz contribution inside the domains. The main
experimental difficulty in determining the DW contribution is to exclude the
other effects. For example in Ref.\cite{gregg} the negative MR observed in
Co films was interpreted in terms of DW scattering. However in Ref.\cite
{ruediger_prb} it was claimed that the predominant contributions to the
observed magnetoresistance of Co films can be explained by a specific
micromagnetic structure, which consists of stripe-domains with magnetization
out-of-the-film plane. In addition, the films show closure caps at the
surfaces with magnetization in plane and parallel to the current. Thus, the
resistivity anisotropy might play a fundamental role. }

{Understanding the details of these experiments is an interesting task.
However, before taking into account all material specific characteristics of
the experiments one should be able to describe general properties of
electron scattering on domain walls. In this paper, we do not try to give an
explanation of all these experiments, but solve an idealized model that may
capture the most general features of transport in the presence of a DW. We
calculate the resistance of a ferromagnetic wire with a DW and restrict
ourself to the case when the magnetization of the ferromagnetic structure
remains always perpendicular to the current. This assumption simplifies the
situation because in this case anisotropic effects do not contribute to the
change of the resistance.}

The DW contribution to the conductance has been considered in several
theoretical works, in which different approaches have been used. For example
in Refs. \cite{tatara,lyanda} quantum effects (weak localization) were taken into
account. It was shown that a DW contributes to the decoherence of electrons
leading to a decrease of the resistance. These effects may be important at
very low temperatures when localization effects start playing a noticeable
role.

At higher temperatures the weak localization effects are not important and
one may try to describe the magnetoresistance in terms of classical motion. In
recent works, Refs. \cite{levy,simanek}, an increase of the resistance due
to a DW was predicted on the basis of a Boltzmann equation. However, the
collision term describing scattering of conduction electrons on impurities
was introduced phenomenologically.

The classical DW resistance was calculated also in Ref. \cite{brataas}. In
that work, it was shown that the DW resistance could be both negative and
positive depending on the difference between the momentum relaxation times $%
\tau _{\uparrow ,\downarrow }$ for the different spin directions. However,
the classical Drude expression for the resistivity was used, in which the
relaxation times $\tau _{\uparrow ,\downarrow }$ were introduced again as
phenomenological parameters. {\ In Refs.\cite{tagirov,bruno} the resistance
of a DW located in a point contact was calculated.}

{\ The purpose of this paper is to derive  a proper
kinetic equation for the distribution function from a microscopic model and to calculate the DW
resistance on its basis. We employ a standard approach based on microscopic
equations for the quasiclassical Green's functions in the Keldysh technique.
Assuming that the impurity scattering potential }$u_{s}${\ is spin dependent,
 we derive the kinetic equation for the matrix (in the Nambu and spin space)
distribution function. As a result we come to the kinetic equation for the
 ditribution function }$f$ that is a $2\times 2$ matrix in the spin space.{\
The impurity scattering potential which enters the collision integral is
also a matrix and this makes the equation considerably more complicated than
the standard one that could be written for a spin independent scattering.}

{\ Throughout this article we assume that the magnetization
remains always perpendicular to the current. First we solve the derived kinetic equation in two simplest cases: a single domain and two domain structure with an abrupt DW (i.e. the width of the DW equals zero).
In the case of a finite DW width
we solve the  kinetic equation assuming  the potentials }$%
u_{\uparrow }${\ and }$u_{\downarrow }${\ do not differ much from each
other. Even in this limit, it is hard to obtain analytical formulae for an
arbitrary width of the DW. Two different limiting cases naturally arise and
this allows us to obtain a solution for the distribution function. The first
limit corresponds to a sharp DW (to a small exchange energy }$J${\ ). The
second limit corresponds to a smooth DW ( to a large }$J${\ ). We note that
only the second limit was analyzed in Refs. \cite{levy,simanek,dugaev}. As
in Refs. \cite{levy,simanek,dugaev} we obtain that the DW increases the
resistance of the system. However our formulae for the contribution of the DW to the
resistance differ essentially from those presented in
Refs. \cite{levy,simanek,dugaev}.}

The paper is organized in the following way. In the next section we
introduce the model and derive the{\ kinetic} equation for the distribution
function in a ferromagnetic wire {\ neglecting quantum effects}. We start
from the microscopic Hamiltonian (\ref{2}) {\ with} different scattering
rates at impurities for spin-up and spin down-electrons. In the subsequent
sections we {\ calculate the conductance of the system} in the diffusive {\
limit}. In section \ref{3a} we consider the case of a sharp DW when $J\ll
D/w^{2}, $ where $D$ is the diffusion coefficient, $w$ is the width of the
DW and $J$ is the exchange field acting on the electron spin. In section \ref{3b}
we calculate the conductance of a ``slowly'' varying DW, {\it i. e.} we
consider the case $D/w^{2}\ll J$. It turns out that in the first case the
conductance is always smaller than in the adiabatic case. In the last
section we summarize our results.

\section{Kinetic Equation}

In this section we derive the kinetic equation for the matrix distribution
function $\hat{f}$ starting from equations for the quasiclassical Green
functions. The function $\hat{f}$ is a $2\times 2$ matrix in the spin space.
We assume that the impurity scattering rate depends on the spin directions
but, for simplicity, we neglect such spin-flip processes as the spin-orbit
interaction or the scattering by magnetic impurities. So, in our model each
impurity scattering vertex is a matrix that does not commute with $\hat{f}$
and therefore the elastic collision integral has a nontrivial form. This
fact has been ignored in Refs. \cite{levy,dugaev}, where the
collision integral was written phenomenologically.

Using the derived kinetic equation we calculate the conductance of a
mesoscopic structure which consists of two reservoirs and a ferromagnetic
wire (or film) connecting the reservoirs (see Fig. 1). A domain wall is
assumed to be present in the ferromagnet. We consider the diffusive limit,
which means that the mean free path is the shortest length (apart from the
Fermi wave length) in the problem. We solve the kinetic equation assuming
the smallness of the parameter $\beta $, defined as

\begin{equation}
\beta =\frac{\sigma _{\uparrow }-\sigma _{\downarrow }}{\sigma _{\uparrow
}+\sigma _{\downarrow }}  \label{1}
\end{equation}
where $\sigma _{\uparrow ,\downarrow }$ are conductivities for different
spin directions. A precise relation between the conductivities $\sigma
_{\uparrow ,\downarrow }$, as well as the diffusion coefficients $%
D_{\downarrow ,\uparrow }$, the corresponding scattering rates will become
clear below.

The assumption $\beta \ll 1$ is valid for ferromagnets with the exchange
energy $J$ much smaller than the Fermi energy. We will consider two limiting
cases: a) $J<<D_{\uparrow ,\downarrow }/w^{2}$ and b) $J>>D_{\uparrow
,\downarrow }/w^{2}$, where $D_{\uparrow ,\downarrow }$ is the diffusion
coefficient for electrons with up and down spins, $w$ is the width of the
DW. The case a) corresponds to a sharp DW. The conductance in this case is
smaller than the conductance of the structure without the domain wall. A
finite width of the domain wall leads to a positive correction to the
conductance. The second case corresponds to a smooth (compared to the
magnetic length $\sqrt{D/J}$ ) DW. In the limit of a large $w$ the
conductance of the structure is close to  that of a structure without a DW.
With decreasing the width of the DW, the conductance of the structure
decreases. { Our results significantly differ significantly from the results obtained in other works, where either the collision term was oversimplified \cite{levy,dugaev}, or the kinetic equation was not treated in a correct way \cite{simanek}}.

We choose the Hamiltonian of the ferromagnet in a simple standard form

\bigskip 
\begin{equation}
H=\sum_{s,s^{\prime }}\int dr\{\psi _{s}^{+}(r)[-\nabla ^{2}/2m+eV(r)-J{\bf n%
}\cdot {\bf \hat{\sigma}}]\psi _{s^{\prime }}(r)\}+H_{imp}\;.  \label{2}
\end{equation}
where $V(r)$ is a smoothly varying (over the wave length $\lambda _{F}$)
electric potential, $J$ is the exchange energy, ${\bf n}$ is the unit vector
directed along the magnetization orientation. The term $H_{imp}(r)$
describes the interaction of electrons with impurities and we assume that it
depends on the spin direction. { The origin of this dependence can be either the band structure or the intrinsec spin dependence of the impurity scattering potential \cite{levy}.} If the magnetization is aligned along the
z-axis, this interaction can be written as

\begin{equation}
H_{imp}=\sum_{i}\int dr\{\psi _{\uparrow }^{+}(r)u_{\uparrow }(r-r_{i})\psi
_{\uparrow}(r)+\psi _{\downarrow}^{+}(r)u_{\downarrow }(r-r_{i})\psi
_{\downarrow}(r)\}\;.  \label{3}
\end{equation}

As in Ref.\cite{BVE_josephson}, we introduce new operators 
\begin{equation}
\psi _{n,s}=\left\{ 
\begin{array}{c}
\psi _{s},n=1 \\ 
\psi _{\overline{s}}^{+},n=2
\end{array}
\right. \;.  \label{a3}
\end{equation}

In terms of the operators $\psi _{n,s}$ and in the case of an arbitrary
angle $\alpha $ between the magnetization vector and the $z$-axis the
Hamiltonian (\ref{3}) can be written as 
\begin{equation}
\begin{array}{lcl}
H_{imp} & = & \sum_{i,n,s}\int dr\psi _{n,s}^{+}(r)\{\hat{\tau}_{3}\otimes 
\hat{\sigma}_{0}u_{+}(r-r_{i})+(\hat{\tau}_{0}\otimes \hat{\sigma}_{3}\cos
\alpha +\hat{\tau}_{3}\otimes \hat{\sigma}_{2}\sin \alpha
)u_{-}(r-r_{i}).\}\psi _{n,s}(r)\} \\ 
& = & \sum_{i,n,s}\int dr\psi _{n,s}^{+}(r)u_{+}\hat{\tau}_{3}\{1+\lambda 
\check{n}\}\psi _{n,s}\;,\label{5}
\end{array}
\end{equation}
where $u_{\pm }=(u_{\uparrow }\pm u_{\downarrow })/2$, $\lambda =u_{-}/u_{+}$
and the matrix $\check{n}$ is defined as $\check{n}=\hat{\tau}_{3}\otimes 
\hat{\sigma}_{3}\exp [-i\alpha \hat{\tau}_{3}\otimes \hat{\sigma}_{1}]$.

Introducing the operators $\psi _{n,s}$, Eq. (\ref{a3}), leads to an
increase of the size of matrix Green functions written below. One has to
deal not only with spin space but also with the Nambu one. Actually, this is
not necessary if one consideres non-superconducting metals only. However,
this extension of the size would become important if the metal wire we
consider were in contact with a superconductor. Although we do not consider
any superconductivity in the present work, we keep at the moment the Nambu
space explicitly having in mind a possible generalization for the
superconductivity.

Now we define the Green functions in the Keldysh technique

\begin{equation}
G_{nn^{\prime }}^{ss^{\prime }}(t_{i},t_{k}^{\prime })=(1/i)<T_{C}(\psi
_{n,s}(t_{i})\psi _{n^{\prime }s^{\prime }}^{+}(t_{k}^{\prime }))>\;.
\end{equation}
where $T_{C}$ means the time ordering along the Keldysh contour $C$. In a
standard way we define the retarded (advanced) $G^{R(A)}$ and Keldysh Green
function $G$ as well as a matrix ${\bf G}$ composed of the matrices $G^{R(A)}
$ and $G$ (see {\it e. g.} Ref.\cite{larkin_ovch_book}). One can obtain an
equation for the matrix ${\bf G}$ in the usual way by summing the ladder
diagrams in the cross technique \cite{AGD} (we neglect all crossed
diagrams). This equation has the form 
\begin{equation}
(i\partial _{t}-H-{\bf \Sigma _{imp}}){\bf G}=1\;  \label{7}
\end{equation}
where $H=[-(1/2m)\nabla ^{2}+V(t)]\hat{\tau}_{3}\otimes \hat{\sigma}_{0}-J%
\hat{\tau}_{3}\mbox{$\check{n}$}$, and the self-energy term ${\bf \Sigma
_{imp}}$ is given by 
\begin{equation}
{\bf \Sigma _{imp}}=n_{imp}\check{u}<{\bf G}>\check{u}\;.  \label{selfenergy}
\end{equation}
Here $\check{u}=u_{+}(1+\lambda \check{n})$, $<{\bf G}>=\nu\int d\xi
_{p}\int d\Omega /4\pi \cdot {\bf G}$ and $n_{imp}$ is the concentration of
impurities. In the quasiclassical approach the density of states $\nu$ is
written in the main approximation with respect to the parameter $J/\epsilon
_{F}$, where $\epsilon _{F}$ is the Fermi energy. In this case $\nu$ is the
same for both spin-up and spin-down electrons. Notice that the r.h.s of Eq. (%
\ref{selfenergy}) is a product of matrices, which in a general case do not
commute.


\begin{figure}
\epsfysize= 8.5cm
\vspace{0.2cm}
\centerline{\epsfbox{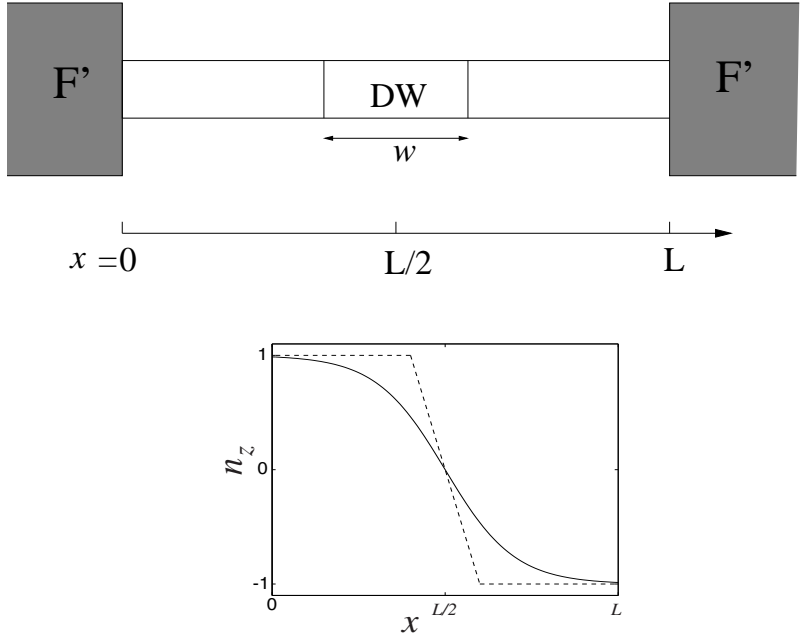}}
\vspace{0.2cm}
\caption{Upper Figure: The geometry considered in this article. The domain
wall (DW) is situated in the middle of the ferromagnetic wire. Lower Figure:
The $z$-component ($n_z(x)$) of the magnetization in the F-wire. The solid
line corresponds to a Bloch-like wall as calculated by Landau and Lifshitz.
The dashed line corresponds to a linear DW}
\label{fig_geom}
\end{figure}

In order to obtain an equation for the quasiclassical Green functions, we
follow the standard way (see for example \cite{larkin_ovch_book} ): we write
the equation conjugate to Eq. (7), multiply both equations by $\hat{\tau}_{3}
$ and subtract from each other. Then, we integrate the final equation over
the variable $\xi _{p}=v_{F}(p-p_{F})$ and obtain

\begin{equation}
\hat{\tau}_{3}\partial _{t}{\bf g}+\partial _{t}{\bf g}\hat{\tau}_{3}+i(eV(t)%
{\bf g}-{\bf g}eV(t))+(v_{F}\nabla ){\bf g}+iJ[\mbox{$\check{n}$},{\bf g}%
]=-(1/2\tau)(\mbox{$\check{m}$}<{\bf g}>\mbox{$\check{m}$}{\bf g}-{\bf g}%
\mbox{$\check{m}$}<{\bf g}>\mbox{$\check{m}$})\;.  \label{8}
\end{equation}
We have introduced the quasiclassical Green function in the usual way 
\begin{equation}
{\bf g}=(i/\pi )\hat{\tau}_{3}\int d\xi _{p}{\bf G}  \label{9}
\end{equation}
The matrix $\mbox{$\check{m}$}$ is equal to $\mbox{$\check{m}$}=1+\lambda %
\mbox{$\check{n}$}$ and $\tau^{-1} =\nu n_{imp}|u_{+}|^{2}$ is the mean momentum
relaxation rate. The elements of the
matrix ${\bf g}$ are $\mbox{$\check{g}$}^{R(A)}$ and $\mbox{$\check{g}$}$

\begin{equation}
{\bf g}=\left( 
\begin{array}{cc}
\mbox{$\check{g}$}^R & \mbox{$\check{g}$} \\ 
0 & \mbox{$\check{g}$}^A
\end{array}
\right)  \label{10}
\end{equation}

Eq.(\ref{8}) is valid in a rather general case. In particular, it can be
employed in the case of a superconductor-ferromagnet structure when the
superconducting condensate penetrates into the ferromagnet. We use Eq.(\ref
{8}) for a normal case, i.e. for F/S structures when one can neglect the
penetration of the condensate into the ferromagnet F or for F/F' structures.
In order to obtain the kinetic equation for the distribution function in the
normal case, we represent the Keldysh component in the usual form \cite
{larkin_ovch_book} 
\begin{equation}
\mbox{$\check{g}$}=\mbox{$\check{g}$}^{R}\cdot \mbox{$\check{f}$}-%
\mbox{$\check{f}$}\cdot \mbox{$\check{g}$}^{A}  \label{11}
\end{equation}
where $\mbox{$\check{g}$}^{R(A)}=\pm \hat{\tau}_{3}\otimes \hat{\sigma}_{0}$
and $\mbox{$\check{f}$}$ is a 4$\times $4 matrix, whose elements are the
components of the distribution function in the Nambu and spin space. This
matrix can be represented in the form 
\begin{equation}
\mbox{$\check{f}$}=\mbox{$\hat{f}$}_{0}\hat{\tau}_{0}+\mbox{$\hat{f}$}_{3}%
\hat{\tau}_{3}\;.  \label{12}
\end{equation}
The components $\mbox{$\hat{f}$}_{0}$ and $\mbox{$\hat{f}$}_{3}$ are
matrices in spin space. In the absence of spin-dependent interactions they
are diagonal and related to the distribution functions for electrons $n$ and
holes $p$ as follows: $\mbox{$\hat{f}$}_{0}=(1-(n_{s}+p_{\overline{s}}))\cdot \hat{%
\sigma}_{0}$; $\mbox{$\hat{f}$}_{3}=-(n_{s}-p_{\overline{s}})\cdot \hat{\sigma}_{0}.$
Taking into account Eqs.(\ref{9}-\ref{12}), one can easily get from Eq.(\ref
{8}) the kinetic equation for the matrix distribution function 
\begin{equation}
\hat{\tau}_{3}(v_{F}\nabla )\mbox{$\check{f}$}+iJ\hat{\tau}_{3}\left[ %
\mbox{$\check{n}$},\mbox{$\check{f}$}\right] =-(1/2\tau)\left[ %
\mbox{$\check{m}$}^{2}\mbox{$\check{f}$}+\mbox{$\check{f}$}\mbox{$\check{m}$}%
^{2}-2\mbox{$\check{m}$}<\mbox{$\check{f}$}>\mbox{$\check{m}$}\right] \;.
\label{13}
\end{equation}
According to all previous definitions we can write 
\begin{equation}
\tau _{\uparrow ,\downarrow }=\frac{\tau}{(1\pm \lambda )^{2}},
\label{14}
\end{equation}
and hence define $\sigma _{\uparrow ,\downarrow }$ and $D_{\uparrow
,\downarrow }$ without using any phenomenological approach. {\ In our model
the conductivities }$\sigma _{\uparrow ,\downarrow }${\ \ are equal to: }$%
\sigma _{\uparrow ,\downarrow }=e^{2}\nu D_{\uparrow ,\downarrow }=e^{2}\nu D(1\pm
\lambda )^{-2},${\ \ where\ }$D=v^{2}\tau/3 $. {\ Note that in the absence of superconductivity the
distribution function has diagonal in the Nambu- space, and therefore one
can take the component (1,1) of Eq. (\ref{13}) and obtain 
\begin{equation}
(v_{F}\nabla )\mbox{$\hat{f}$}+iJ\left[ \mbox{$\hat{n}$},%
\mbox{$\hat{f}$}\right] =-(1/2\tau)\left[ \mbox{$\hat{m}$}^{2}\mbox{$\hat{f}$}%
+\mbox{$\hat{f}$}\mbox{$\hat{m}$}^{2}-2\mbox{$\hat{m}$}<\mbox{$\hat{f}$}>%
\mbox{$\hat{m}$}\right] \;,  \label{spin}
\end{equation}
where all matrices are now} $2\times2$  matrices in the spin
space. In particular, $\mbox{$\hat{m}$}=1+\lambda \mbox{$\hat{n}$}$ and $%
\mbox{$\hat{n}$}=\hat{\sigma}_{3}\exp [-i\alpha \hat{\sigma}_{1}]$. { Note that the left-hand side of Eq. (\ref{spin}) coincides with  the left-hand side of the well known kinetic equation derived for a magnetic material (see for example Ref. \cite{abrikosov}, where the kinetic equation is presented for a dynamic case in the absence of scattering by impurities)}. The
solution $\hat{f}$ of this equation coincides with the component (1,1) of
the distribution function $\check{f}$, which satisfies Eq. (\ref{13}). Since
in this article normal materials (no superconductors) are considered, we will
analyze Eq. (\ref{spin}). { One can exclude the spatial dependence of the matrices $\hat{n}$ and $\hat{m}$ performing an unitary transformation defined by
\[
\hat{f}=\hat{U}\cdot \hat{\tilde{f}}\cdot \hat{U}^{+},\; \; 
\hat{U}= \hat{\sigma} _{0}\cos \alpha /2+i \hat{\sigma} _{1}\sin \alpha /2\; .
\]
In this  case one obtains an equation for the distribution function $\hat{\tilde{f}}$}
{
\begin{equation}
(v_{F}\nabla )\mbox{$\hat{\tilde{f}}$}+ i(v_F/2)\alpha'(x)\left[\hat{\sigma}_1,
\mbox{$\hat{\tilde{f}}$}\right] + iJ\left[ \mbox{$\hat{\sigma}_3$},%
\mbox{$\hat{\tilde{f}}$}\right] =-(1/\tau)\left[ \mbox{$\hat{\tilde{f}}$}-<\mbox{$\hat{\tilde{f}}$}>+\lambda\left[\hat{\sigma}_3,\mbox{$\hat{\tilde{f}}$}-<\mbox{$\hat{\tilde{f}}$}>\right]+\lambda^2\left(\mbox{$\hat{\tilde{f}}$}-\hat{\sigma}_3<\mbox{$\hat{\tilde{f}}$}>\hat{\sigma}_3\right)\right] \;.\label{rotated_kin}
\end{equation}
The left-hand side of this equation differs from the one derived in Ref.\cite{simanek}. In the latter there is an additional term of the form $(\alpha'(x)/4m)[\hat{\sigma}_1,(\partial F/\partial x)]_+$ which, as we have shown, does not appear in the quasiclassical approach. Moreover, due to this term the kinetic equation of Ref. \cite{simanek} violates the particle number conservation and therefore leads to wrong results. Notice also,  that the collision term ( right-hand side of Eq. (\ref{rotated_kin}) ) after the unitary rotation may not be diagonal in spin space. This fact was ignored in Refs. \cite{levy,dugaev}. We will see in the next sections that in the case $(D/w^2)\ll J$, it is convenient to work with Eq.(\ref{rotated_kin}), while in the opposite case it is easier to solve the kinetic equation in its original form Eq. (\ref{spin}).
}
We assume that the system is diffusive (this implies the condition $J\tau <<1
$){\ . In this case one can expand the distribution function }$\hat{f}$ {\ \
in spherical harmonics and consider only the first two of them} 
\begin{equation}
\mbox{$\hat{f}$}=\mbox{$\hat{s}$}+\mu \mbox{$\hat{a}$}\;,  \label{sphharm}
\end{equation}
{\ where }$\mu =\cos \theta ${\ \ and }$\theta ${\ \ is the angle between }$%
v_{F}${\ \ and the }$x${\ -axis. Using Eqs. (\ref{spin}) and (\ref{sphharm}%
), one obtains two equations for the functions }$\mbox{$\hat{s}$}${\ \ and }$%
\mbox{$\hat{a}$}$ 
\begin{eqnarray}
v_{F}\partial _{x}\mbox{$\hat{s}$}+iJ\left[ \mbox{$\hat{n}$},\mbox{$\hat{a}$}%
\right]  &=&-(1/2\tau)\left( \mbox{$\hat{m}$}^{2}\mbox{$\hat{a}$}+%
\mbox{$\hat{a}$}\mbox{$\hat{m}$}^{2}\right)   \label{e1} \\
(v_{F}/3)\partial _{x}\mbox{$\hat{a}$}+iJ\left[ \mbox{$\hat{n}$},%
\mbox{$\hat{s}$}\right]  &=&-(1/2\tau)\left( \mbox{$\hat{m}$}^{2}%
\mbox{$\hat{s}$}+\mbox{$\hat{s}$}\mbox{$\hat{m}$}^{2}-2\mbox{$\hat{m}$}%
\mbox{$\hat{s}$}\mbox{$\hat{m}$}\right) \;.  \label{e2}
\end{eqnarray}
{\ In the second equation we have performed an averaging over the angle }$%
\mu ${\ . The boundary conditions at the interfaces with the reservoirs are
given by imposing the continuity of the symmetric part }$\mbox{$\hat{s}$}(x)$%
{\ \ of the distribution function (we assume a perfect contact of the F wire
with the reservoirs)} 
\begin{equation}
\mbox{$\hat{s}$}(L)=\tanh \frac{\epsilon }{2T}\hat{\sigma}_{0}\;  \label{bc1}
\end{equation}
and 
\begin{equation}
\mbox{$\hat{s}$}(0)=\tanh \frac{\epsilon +eV}{2T}\hat{\sigma}_{0}\;.
\label{bc2}
\end{equation}
{\ Once we determine the distribution function $\mbox{$\hat{f}$}$, we can
calculate the current density using the following expression 
\begin{equation}
j=-\frac{1}{4}e\nu\frac{v_{F}}{3}\int d\epsilon {\rm Tr}\mbox{$\hat{a}$}\;.
\label{current}
\end{equation}
}

{\ In the next sections we determine the resistance of a domain wall with a
finite width. Here on the basis of Eq.(\ref{spin}), the conductance of a
F'/F/F' mesoscopic system is calculated in the simplest cases: a single
domain in the ferromagnetic wire and a two-domain structure in the F wire
with an abrupt domain wall (i.e., }$w=0,${\ \ see Fig.1). In this case} (the
magnetization is parallel or antiparallel to the z-axis), both parts of the
distribution function $\mbox{$\hat{s}$}$ {\ and }$\mbox{$\hat{a}$}$ {\ are
proportional to } $\hat{\sigma}_{0,3}$. {\ Therefore the commutator on the
left hand side is equal to zero.} {\ From Eq. (\ref{e1}) we find }

\begin{equation}
\mbox{$\hat{a}$}=-\mbox{$\hat{m}$}^{-2}v_{F}\partial _{x}\mbox{$\hat{s}$}\;.
\label{a1d}
\end{equation}
{\ We substitute this expression into equation (\ref{e2}). Taking into
account that the right-hand side is zero, we obtain after integration} 
\begin{equation}
\mbox{$\hat{s}$}=\mbox{$\hat{s}$}(0)+\mbox{$\hat{m}$}^{2}\hat{I}x/D\;,
\label{s1d}
\end{equation}
{\ The integration constant or, in other words, the ``partial current'' per
unit energy }$I${\ \ is found from the boundary condition (\ref{bc2})} 
\begin{equation}
\mbox{$\hat{I}$}=-\frac{D}{L}\mbox{$\hat{m}$}^{-2}F_{-}\hat{\sigma}_{0}\;,
\label{I1}
\end{equation}
{\ where }$\mbox{$\hat{m}$}^{-2}=[1+\lambda ^{2}-2\lambda \mbox{$\hat{n}$}%
]/(1-\lambda ^{2})$ and $F_{-}=\tanh \frac{\epsilon +eV}{2T}-\tanh \frac{%
\epsilon }{2T}$. {\ \ Substituting this expression into Eq. (\ref{current}),
we find the current and the differential conductance }$G=\left. dI/dV\right|
_{V=0}$ 
\begin{equation}
G_{1d}=(2\sigma /L)(1+\lambda ^{2})/(1-\lambda^2)^{2}=G_{\uparrow
}+G_{\downarrow }  \label{G1d}
\end{equation}
{\ Here }$G_{\uparrow ,\downarrow }=\sigma _{\uparrow ,\downarrow }/L$, and $%
\sigma =e^{2}\nu D$.{\ \ Thus, the conductance has the usual form. We note that
in terms of $\lambda $ the conductivities $\sigma _{\uparrow ,\downarrow }$
are given by $\sigma _{\uparrow ,\downarrow }=\sigma /(1\pm \lambda )^{2}$,
and hence the coefficient $\beta $ defined in Eq. (\ref{1}) is related to $%
\lambda $ via the relation $\beta =-2\lambda /(1+\lambda ^{2})$.}

{\ Let us consider the same system with two domains in the F wire and with
an abrupt DW located in the middle of the wire. In this case }$\alpha =0$ {\
in the interval} $0<x<L/2$ and {\ \ }$\alpha =\pi $ {\ in the interval} $%
L/2<x<L.$ Eqs.{\ (\ref{e1}-\ref{e2}) are solved in the same way as for the
single domain case. For the symmetric part of the distribution function we
obtain} 
\begin{equation}
\mbox{$\hat{s}(x)$}=\left\{ 
\begin{array}{cr}
\hat{\sigma}_{0}\tanh \frac{\epsilon +eV}{2T}+\mbox{$\hat{m}$}^{2}(0)%
\mbox{$\hat{I}$}x/D & {\ 0<x<L/2} \\ 
s(L/2)+\mbox{$\hat{m}$}^{2}(\pi )\mbox{$\hat{I}$}(x-L/2)/D & {\ L/2<x<L}
\end{array}
\right. \;,  \label{s2d}
\end{equation}
{\ where }$\mbox{$\hat{m}$}^{2}(0)=\mbox{$\hat{m}$}^{2}|_{\alpha =0}${\ .
The integration constant again is found from the boundary condition (\ref
{bc2}). We get for }$\mbox{$\hat{I}$}$ 
\begin{equation}
\mbox{$\hat{I}$}=\hat{\sigma}_{0}DF_{-}/(1+\lambda ^{2})L  \label{I2d}
\end{equation}
{\ and for the conductance} 
\begin{equation}
G_{2d}=(2\sigma /L)/(1+\lambda ^{2})=4G_{\uparrow }G_{\downarrow
}/(G_{\uparrow }+G_{\downarrow })  \label{G2d}
\end{equation}
{This result has been obtained earlier (see Ref. \cite{ebels} and
references therein). In the next section we calculate }$G$ f{or the case
when the magnetization (or the vector ${\bf n}$) rotates in the y-z plane
over a finite length }$w$.

\section{ Conductance of a domain wall}

{\ The problem of calculating the conductance for a system with a finite
width of a DW is rather complicated. In order to simplify it, we make an
assumption that the scattering times }$\tau _{\uparrow ,\downarrow }${\ \
are close to each other, i.e.}

\begin{equation}
\lambda \ll 1\;.  \label{lambda}
\end{equation}
This condition is met in ferromagnets with an exchange energy $J$ smaller
than the Fermi energy. We consider again the system shown in Fig. \ref
{fig_geom}. The total length of the ferromagnetic wire is $L$. A Bloch-like
DW is situated in the region $(L-w)/2<x<(L+w)/2$ and separates two domains
with opposite magnetizations. Thus, the effective width of the DW is $w$. 
It is not easy to obtain the exact solution of Eqs.(\ref{e1},\ref{e2}).
However one can assume that condition (\ref{lambda}) is satisfied and expand
the functions $\mbox{$\hat{s}$}$ and $\mbox{$\hat{a}$}$ up to terms
proportional to $\lambda ^{2}$. We distinguish two cases: a) $J\ll D/w^{2}$,
which corresponds to a sharp DW; and b) $J\gg D/w^{2}$.

\subsection{Small exchange energy}\label{3a}

{\ If the exchange field is weak (}$J\ll D/w^{2}${\ ) or the DW wall is very
sharp, one can easily solve Eqs. (\ref{e1},\ref{e2}) for an arbitrary form
of the DW. We assume that the DW width exceeds the mean free path but is
smaller than the magnetic length }$\xi _{J}=\sqrt{D/J}${\ . In this case, we
expand the solution of Eqs. (\ref{e1},\ref{e2}) in the small parameters }$%
Jw^{2}/D${\ \ and }$\lambda ${\ . In the zero order approximation, we get}

\begin{equation}
\mbox{$\hat{a}$}_{0}=-l\partial _{x}\mbox{$\hat{s}$}_{0}  \label{a0}
\end{equation}
and 
\begin{equation}
D\partial _{x}\mbox{$\hat{s}$}_{0}=\hat{I}_{0};\text{ \ \ \ \ }%
\mbox{$\hat{s}$}_{0}=\hat{\sigma}_{0}\tanh \frac{\epsilon +eV}{2T}+%
\mbox{$\hat{I}$}_{0}x/D\;,  \label{s0}
\end{equation}
{\ where the `partial current' is found from the boundary condition (\ref
{bc2}) and is equal to } 
\begin{equation}
\hat{I}_{0}=-\hat{\sigma}_{0}DF_{-}/L  \label{I0}
\end{equation}
{\ In the first approximation we find from Eq.(\ref{e1})} 
\begin{equation}
\mbox{$\hat{a}$}_{1}=-l\partial _{x}\mbox{$\hat{s}$}_{1}-2\lambda %
\mbox{$\hat{n}$}(x)\mbox{$\hat{a}$}_{0}  \label{a1}
\end{equation}
{\ The solution of Eq. (\ref{e2}) for the symmetric part }$\mbox{$\hat{s}$}%
_{1}${\ \ has the form} 
\begin{equation}
\mbox{$\hat{s}$}_{1}=\hat{I}_{1}x/D+2\lambda \hat{I}_{0}\int_{0}^{x}dx_{1}%
\mbox{$\hat{n}$}(x_{1})/D  \label{s1}
\end{equation}
{\ This and next corrections should satisfy zero boundary conditions.
Therefore we find for }$\hat{I}_{1}$ 
\begin{equation}
\hat{I}_{1}=-2\lambda \hat{I}_{0}<\mbox{$\hat{n}$}>_{L}
\end{equation}

{where }$<(...)>_{L}=1/L\int_{0}^{L}(...)dx.${\ \ As it follows from Eq.(\ref
{current}), the first correction does not contribute to the current. The
zero order correction leads to the expression for the conductance given by
Eq. (\ref{G2d}) if we expand it in the small parameter }$\lambda ${\ \ (the
case of an abrupt DW). In order to find a correction to the conductance due
to a finite width of the DW, one has to find the second order corrections.
One can see from Eq.(\ref{current}) that only components of }$%
\mbox{$\hat{a}$}_{2}${\ \ or }$\mbox{$\hat{s}$}_{2}$ {\ proportional to }$%
\hat{\sigma}_{0}${\ \ contribute to the current. Therefore we take the trace
in the spin space from Eq.(\ref{e1}) and Eq.(\ref{e2}) and find easily}

\begin{equation}
\mbox{${\rm Tr}$}\mbox{$\hat{a}$}_{2}=-(l/D)\mbox{${\rm Tr}$}\hat{I}_{2}
\end{equation}
and 
\begin{equation}
\mbox{${\rm Tr}$} \mbox{$\hat{s}$}_{2}=-(2\lambda )^{2}\mbox{${\rm Tr}$}\hat{%
I}_{0}<\mbox{$\hat{n}$}>_{L}\left[\int_{0}^{x}dx_{1}\mbox{$\hat{n}$}(x_{1})-<%
\mbox{$\check{n}$}>x\right]/D\; ,  \label{s2}
\end{equation}
where 
\begin{equation}
\mbox{${\rm Tr}$}\hat{I}_{2}=-(1/2)\mbox{${\rm Tr}$}_{\hat{\sigma}}\hat{I}%
_{0}\lambda ^{2}\left[1-4<\mbox{$\hat{n}$}>_{L}^{2}\right]\; .  \label{I2}
\end{equation}
{\ Using Eqs.(\ref{I0}) and Eqs.(\ref{I2}) we obtain the expression for the
conductance which can be represented in the form} 
\begin{equation}
G=G_{2d}\left[1+(2\lambda )^{2}\frac{1}{2}\mbox{${\rm Tr}$}\,<%
\mbox{$\hat{n}$}>_{L}^{2}\right]  \label{G2}
\end{equation}
{\ This formula determines the conductance of the system under consideration
for the case when the precession frequency }$J$ {\ is smaller than the
inverse time of diffusion of an electron through the DW. One can see that in
the case of a DW with a finite width the conductance is larger than in the
case of a sharp DW ({\it cf.} Eq. (\ref{G2d})), but smaller than the
conductance in the single domain case ({\it cf.} Eq. (\ref{G1d})) }. Note that Eq. (\ref{G2}) has been obtained in the limit of small $\lambda$. Therefore the conductance $G_{2d}$ should be expanded in $\lambda$ (see Eq. (\ref{G2d})) and terms of order higher than $\lambda^2$ should be neglected. { There is an interesting consequence from the result of Eq. (\ref{G2}). Let us consider the case of two DWs separating three regions of length $d$ with homogeneous magnetization. For simplicity we assume that the shapes of the DWs are described by a piece-wise linear function, which is characterized by a wave vector ${\bf Q}=(w/\pi,0,0)$. If one defines the chirality vector as $\delta {\bf v_{ch}}={\bf n}(x)\times{\bf n}(x+\delta x)$, where ${\bf n}(x)$ is the unit vector directed along the local magnetization, two cases should be distinguished:
a) the DWs have different chirality. In this case ${\rm Tr}<\hat{n}>^2_L=(2/L^2)(d^2+16(w^2/\pi^2))$. Thus we see that an additional DW will decreases the conductance of the system.
b) the chiriality vectors have different signs. In this case  ${\rm Tr}<\hat{n}>^2_L=(2/L^2)d^2$, and hence the contributions of both DWs to the conductance cancel each other. 
This result can be generalized easily for an arbitrary number of DWs
}

\begin{figure}
\epsfysize= 8cm
\vspace{0.2cm}
\centerline{\epsfbox{ 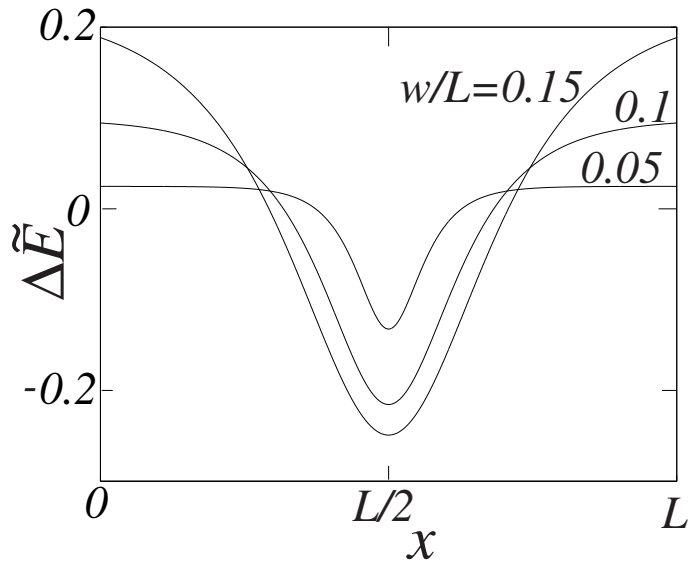}}
\vspace{0.2cm}
\caption{The spatial distribution of the electrical field in the F-wire for
different values of $w/L$. Here $\Delta\tilde{E}=(E(x)-E_0)/(2\protect\lambda%
)^2$.}
\label{fig_2}
\end{figure}

Now we calculate the spatial distribution of the electric field in the
ferromagnetic wire showed in Fig. \ref{fig_geom}. The electric potential $V(x)${\ \ is given by the
expression (see, for example, Ref. }\cite{larkin_ovch_book}) 
\begin{equation}
V(x)=(1/4)Tr \hat{\sigma} _{0}\int d\epsilon \mbox{$\hat{s}$}
\label{potential}
\end{equation}
{\ According to Eqs}.{\ (\ref{s0}) and (\ref{s2}) the electric field }$%
E(x)=-\partial _{x}V(x)$ {\ in the ferromagnetic wire is given by} 
\begin{equation}
E(x)=(V/L)\{1+(2\lambda )^{2}[<\cos \alpha >_{L}(<\cos \alpha >_{L}-\cos
\alpha )+<\sin \alpha >_{L}(<\sin \alpha >_{L}-\sin \alpha )]\}\; .
\label{E(x)}
\end{equation}
{\ For example, if we consider the structure of the Bloch wall which has
been calculated by Landau and Lifshitz \cite{landau_wall} 
\begin{equation}
\cos \alpha =\tanh [(x-L/2)/w];\text{ \ \ }\sin \alpha =\cosh
^{-1}[(x-L/2)/w]  \label{Landau}
\end{equation}
we obtain 
\begin{equation}
E(x)-E_{0}=E_{0}(2\lambda )^{2}(\pi w/L)\left\{(\pi w/L)-\cosh
^{-1}[(x-L/2)/w]\right\}\; ,  \label{E(x)L}
\end{equation}
where $E_{0}=(V/L).$ In Fig.\ref{fig_2} we plot the dependence $E(x)$ given
by Eq. {\ (\ref{E(x)L}). One can see that in the region of the DW the
electric field decreases; this means an increase in the local conductivity.} 
}

{\ \ In the next section we consider the case of a strong exchange field or
of a wide wall, i.e. the case }$w\gg \xi _{J}${\ .}


\subsection{ Large Exchange Energy}\label{3b}

Now we consider the case with a large exchange energy or a slow variation of
the direction of the magnetization within the DW ($w\gg \sqrt{D/J}$). In
this case, the problem becomes more complicated because we cannot neglect
the commutator on the left hand side in Eqs. (\ref{e1},\ref{e2}) and cannot
find a solution of these equations even for the case of small $\lambda $.
Therefore we simplify the problem assuming that the shape of the DW is
described by a piece-wise linear function (see Fig. \ref{fig_geom})

\bigskip 
\begin{equation}
\mbox{$\hat{n}$}=\left\{ 
\begin{array}{cr}
\hat{\sigma} _{3} & {\rm in\;region\;I} \\ 
\hat{\sigma} _{3}\exp [-i(\pi/w)(x-(L-w)/2)\hat{\sigma} _{1}] & {\rm within\;the\;DW}
\\ 
-\hat{\sigma} _{3} & {\rm in\;region\;III.}
\end{array}
\right.  \label{n}
\end{equation}

Obviously the results for other shapes of the DW like that given by Eq. (\ref
{Landau}) will differ from ours only by a numerical factor. We again expand
the solution in the small parameter $\lambda $, {\it i.e.} $\mbox{$\hat{a}$}=%
\mbox{$\hat{a}$}_{0}+\mbox{$\hat{a}$}_{1}+\mbox{$\hat{a}$}_{2}+...$. The
zeroth order terms can be obtained easily as before and they are given by
Eqs.(\ref{a0},\ref{s0}). The first correction $\mbox{$\hat{a}$}_{1}$ is
given again by Eq. (\ref{a1}) and the first correction for the symmetric
part $\mbox{$\hat{s}$}_{1}$ obeys the equation

\begin{equation}
D\partial _{xx}^{2}\mbox{$\hat{s}$}_{1}-iJ[\mbox{$\hat{n}$}%
(x),s_{1}]=2\lambda \hat{I}_{0}\partial _{x}\mbox{$\hat{n}$}\;.  \label{Ls1}
\end{equation}
This equation can be solved for the case of $\mbox{$\hat{n}$}(x)$ given by
Eq.(\ref{n}) with the help of an unitary transformation (a rotation in the
spin space). We do not need to find the second order corrections $%
\mbox{$\hat{a}$}_{2}$ and $\mbox{$\hat{s}$}_{2}$, since the sought
correction to the conductance can be expressed in terms of $\mbox{$\hat{s}$}%
_{1}.$ Indeed, let us write the equation for $\mbox{${\rm Tr}$}%
\mbox{$\hat{s}$}_{2}$ which follows from Eq.(\ref{e2}) 
\begin{equation}
\mbox{${\rm Tr}$}\{D\partial _{x}\mbox{$\hat{s}$}_{2}+(v_{F}/3)[\lambda ^{2}\mbox{$\hat{a}$}_{0}+2\lambda \mbox{$\hat{n}$}%
\mbox{$\hat{a}$}_{1}]-\mbox{$\hat{I}$}_{2}\}=0\;,  \label{Ls2}
\end{equation}
where $\mbox{$\hat{I}$}_{2}$ is the integration constant which is related to 
$\mbox{$\hat{a}$}_{2}:$ $\mbox{${\rm Tr}$}\hat{\sigma}_{0}\mbox{$\hat{a}$}%
_{2}=-l\mbox{$\hat{I}$}_{2}/D$. We integrate this equation from $0$ to $L$
taking into account the boundary conditions at $x=0$ and $x=L$: $%
\mbox{$\hat{s}$}_{2}=0.$ After simple transformations we obtain

\begin{equation}
\mbox{${\rm Tr}$}\{\mbox{$\hat{I}$}_{2}-3\lambda ^{2}\mbox{$\hat{I}$}%
_{0}\}/D=2\lambda \mbox{${\rm Tr}$}\hat{\sigma} _{0}\{\int_{0}^{L}dx%
\mbox{$\hat{s}$}_{1}(x)\mbox{$\hat{n}$}(x)\}; .  \label{LI2}
\end{equation}
As before, the correction to the conductance is determined by $%
\mbox{$\hat{I}$}_{2}.$ We see that in order to find this correction, one has
to solve Eq. (\ref{Ls1}) for $\mbox{$\hat{s}$}_{1}.$ This equation can be
solved with the help of the unitary transformation

\begin{equation}
\mbox{$\hat{s}$}_{1}=\hat{U}\cdot \mbox{$\hat{S}$}\cdot \hat{U}^{+},\; \; 
\hat{U}= \hat{\sigma} _{0}\cos \alpha /2+i \hat{\sigma} _{1}\sin \alpha /2%
\text{\ \ \ }  \label{U-Trans}
\end{equation}

\begin{figure}
\epsfysize= 8cm
\vspace{0.2cm}
\centerline{\epsfbox{ 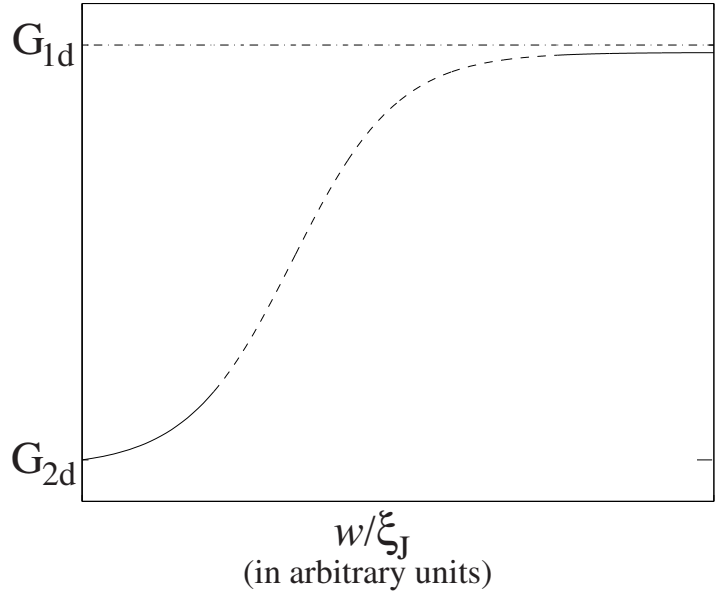}}
\vspace{0.2cm}
\caption{Schematical representation of the conductance as a function of the
width $w$ of the DW. In the intermediate region (dashed line) the curve is
extrapolated from our results. }
\label{fig_3}
\end{figure}

This rotation transforms the vector $\mbox{$\hat{n}$}$ into $\mbox{$\hat{N}$}%
= \hat{\sigma} _{3}.$ Performing the $U$-transformation, we obtain instead
of Eq.(\ref{LI2})

\begin{equation}
\mbox{${\rm Tr}$}\hat{\sigma} _{0}\{\mbox{$\hat{I}$}_{2}-3\lambda ^{2}%
\mbox{$\hat{I}$}_{0}\}/D=2\lambda \partial _{x}\alpha \mbox{${\rm Tr}$}_{%
\hat{\sigma}}\hat{\sigma} _{2}<\mbox{$\hat{S}$}(x)>\; .  \label{UI2}
\end{equation}
After the $U$-transformation Eq.(\ref{Ls1}) acquires the form (in the region
of the DW)

\begin{equation}
\partial _{xx}^{2}\mbox{$\hat{S}$}-(Q^{2}/2)(\mbox{$\hat{S}$}-\hat{\sigma}%
_{1}\mbox{$\hat{S}$}\hat{\sigma}_{1})+iQ(\hat{\sigma}_{1}\partial _{x}%
\mbox{$\hat{S}$}-\partial _{x}\mbox{$\hat{S}$}\hat{\sigma}_{1})-iJ[%
\mbox{$\hat{N}$},\mbox{$\hat{S}$}]=2\lambda \hat{\sigma}_{2}\mbox{$\hat{I}$}%
_{0}Q/D\;,  \label{LS}
\end{equation}
where $Q=\partial _{x}\alpha $ and $\mbox{$\hat{N}$}=\hat{\sigma}_{3}$. This
equation is valid in the region of the DW, whereas in the regions I and III
we have to set $Q=0$ and to take into account that in the region III $%
\mbox{$\hat{N}$}=-\hat{\sigma}_{3}.$ The matrix $\mbox{$\hat{S}$}$ should be
represented as a sum: $\mbox{$\hat{S}$}=S_{1}\hat{\sigma}_{1}+S_{2}\hat{%
\sigma}_{2}+S_{3}\hat{\sigma}_{3}.$ The components $S_{k}$ are given by a
linear combination of the eigen-functions of Eq. (\ref{LS}). They obey zero
boundary conditions at $x=0$ and $x=L$ and should be matched at $%
x=L_{1}=(L-w)/2$ and $x=L_{2}=(L+w)/2$. The eigen-values of Eq.(\ref{LS}) ($%
S_{k}\sim \exp (\kappa x))$ are determined by the equation

\begin{equation}
\kappa ^{2}(\kappa ^{2}+Q^{2})^{2}=(\kappa ^{2}-Q^{2})/\xi _{J}^{2}\; ,
\label{kappa}
\end{equation}
where $\xi _{J}^{-2}=2J/D.$ In a general case a solution of Eq. (\ref{LS})
has a cumbersome form. We represent here the form of a solution for $%
\mbox{${\rm Tr}$}\hat{\sigma} _{2}\mbox{$\hat{S}$}(x)$ in the region of the
DW which we are interested in:

\begin{equation}
\mbox{${\rm Tr}$}\hat{\sigma} _{2}\mbox{$\hat{S}$}(x)\cong -2\lambda%
\mbox{${\rm Tr}$}_{\hat{\sigma}} \mbox{$\hat{I}$}_{0}(\xi _{J}^{2}Q/D)%
\mathop{\rm Im}%
[\exp (-(1+i)(x-L_{1})/\xi _{J}\text{\ }\sqrt{2})\text{\ }+\exp
((1+i)(x-L_{2})/\xi _{J}\text{\ }\sqrt{2})]\; .  \label{S(x)}
\end{equation}
We dropped terms of the higher order in the parameter $Q\xi _{J}$\ $\sim \xi
_{J}/w.$ Using this expression and Eq. (\ref{UI2}), we readily get the
expression for the current and for the conductance $G$

\begin{equation}
G=G_{1d}\left( 1-\frac{\pi ^{2}\xi _{J}^{3}}{Lw^{2}}\lambda ^{2}\right) \;,
\label{cond2}
\end{equation}
where $G_{1d}$ is the conductance for a homogenous magnetized wire (see Eq. (%
\ref{G1d})). Again terms of order higher than $\lambda^2$ should be neglected. Note that $G_{1d}$ is always larger than the conductance in the
case of a two domain wire $G_{2d}$ (see Eq. (\ref{G2d})). Eq. (\ref{cond2})
shows that the DW decreases the conductance compared to the conductance $%
G_{1d}$ of a single domain F wire. Our result is sketched in Fig. \ref{fig_3}%
. We see that within our approach a DW with a finite width is always a source of resistance.

\section{Conclusion}

Using a simple microscopic model (equal density-of-states but different
impurity scattering times $\tau _{\uparrow ,\downarrow }$ for electrons with
spin up and down), we have derived the kinetic equation for the matrix
distribution function. The derivation has been performed by a standard
method on the basis of microscopic equations for the quasiclassical Green
functions in the Keldysh technique. This equation can be applied to the
studies of transport in, for example, ferromagnets with a non-homogeneous
magnetization. 

We have employed this equation to calculate the conductance $G$ in a
mesoscopic F'/F/F' structure. We have assumed that the parameter $\lambda
=(\tau _{\downarrow }-\tau _{\uparrow })/$ $2(\tau _{\downarrow }+\tau
_{\uparrow })$ is small and the length of the F wire $L$ is shorter than the
spin energy relaxation length. Two different limits appear which are
determined by the product of the exchange energy $J$ and the diffusion
time $\tau _{w}=w^{2}/D$ of electrons through the DW. In the limit $\tau
_{w}J\ll 1$ and a very thin DW the conductance of the structure (per the
unit cross-section area) is equal to $G_{2d}=4\sigma _{\uparrow }\sigma
_{\downarrow }(\sigma _{\uparrow }+\sigma _{\downarrow })/L$. The account
for a finite width of the DW leads to an increase in the conductance by a
normalized amount of the order ($\lambda w/L)^{2}.$ We have also calculated
in this limit the spatial distribution of the electric field in the F wire.
The electric field has a minimum in the center of the DW which corresponds
to an enhanced local conductivity. In the other limit $\tau _{w}J>>1$
(adiabatic variation of the magnetization in the DW) the conductance
coincides in the main approximation with that of a single domain structure $%
G_{1d}=(\sigma _{\uparrow }+\sigma _{\downarrow })/L$. The account for
rotation of the magnetization in the DW leads to a negative correction to
the conductance of the order -$\lambda ^{2}(\tau _{w}J)^{-3/2}(w/L).$ Our
results differ from those published earlier \cite{levy,simanek,dugaev}
because in the latter works the collision term was written
phenomenologically. In particular the matrix character of the impurity
vertex was not taken into account.

We would like to thank SFB 491 for financial support.



\end{document}